# Ultra fast quantum key distribution over a 97 km installed telecom fiber with wavelength division multiplexing clock synchronization

Akihiro Tanaka,[1,*] Mikio Fujiwara,[2] Sae Woo Nam,[3] Yoshihiro Nambu,[4] Seigo Takahashi,[1] Wakako Maeda,[1] Ken-ichiro Yoshino,[4] Shigehito Miki,[5] Burm Baek,[3] Zhen Wang,[5] Akio Tajima,[1] Masahide Sasaki,[2] and Akihisa Tomita[4,6]

[1] *System Platforms Research Laboratories, NEC Corporation, 1753, Shimonumabe, Nakahara-ku, Kawasaki, Kanagawa 211-8666, Japan*
[2] *New Generation Network Research Center, National Institute of Information and Communications Technology, 4-2-1 Nukui-Kitamachi, Koganei, Tokyo 184-8795, Japan*
[3] *National Institute of Standards and Technology, 325 Broadway, Boulder, Colorado 80305, USA*
[4] *Nano Electronics Research Laboratories, NEC Corporation, 34 Miyukigaoka, Tsukuba, Ibaraki 305-8501, Japan*
[5] *Kobe Advanced ICT Research Center, National Institute of Information and Communications Technology, 588-2, Iwaoka, Nishi-ku, Kobe, Hyogo 651-2492, Japan*
[6] *Quantum Computation and Information Project, JST ERATO-SORST, 5-28-3, Hongo, Bunkyo-ku, Tokyo 133-0033 Japan*
[*]*Corresponding author: a-tanaka@dh.jp.nec.com*

**Abstract:** We demonstrated ultra fast BB84 quantum key distribution (QKD) transmission at 625 MHz clock rate through a 97 km field-installed fiber using practical clock synchronization based on wavelength-division multiplexing (WDM). We succeeded in over-one-hour stable key generation at a high sifted key rate of 2.4 kbps and a low quantum bit error rate (QBER) of 2.9%. The asymptotic secure key rate was estimated to be 0.78-0.82 kbps from the transmission data with the decoy method of average photon numbers 0, 0.15, and 0.4 photons/pulse.

**OCIS codes:** (270.5568) Quantum cryptography; (060.5565) Quantum communication

**1. Introduction**

Secure cryptographic key against all possible attacks can be shared with some QKD protocols such as Bennett-Brassard 1984 (BB84) [1]. The unconditional security for BB84 has been proven even with an attenuated laser source [2-8]. Demonstrated transmission performance is, however, still far from practical use. The secure key rate for BB84 over 100 km remains around 10 bps [9, 10]. Although a higher key rate about 17 kbps was marked over 100 km with differential phase shift QKD using superconducting single photon detectors (SSPDs) [11], its security proof is yet restricted to a limited class of attacks. Moreover, these demonstration experiments were done in laboratories with clock synchronization through short electric cables, which cannot be applied to a field environment experiment.

In this paper, we report a QKD field test through a 97 km installed fiber with the repetition rate of 625 MHz, using one-way interferometers based on planar light circuit (PLC), SSPDs and timing information delivery using a classical signal transmitted with the quantum signal in the fiber. We establish the clock transmission technique for long distance QKD without degrading the quantum signal, using an optical amplifier for the clock signal, narrowband filters on both quantum and clock signals. Over-one-hour stable key generation with an untrampled high sifted key rate of 2.4 kbps and considerably low QBER of 2.9% after 97 km transmission is achieved. To our knowledge, this result is the world's fastest BB84 QKD field test through a 97 km or longer optical fiber. An ultra-high speed secure QKD system will be completed by combining data acquisition, control, and key distillation electronics with the present transmission system.

**2. High speed, long distance and long term QKD field test**

*2.1 High speed and long distance QKD*

QKD has recently been put to practical demonstration through field-installed fibers [12-15]. To overcome environmentally dependent fluctuations of fiber transmission properties, a bidirectional transmission (plug and play) scheme was designed [16] and applied to field QKD experiments over a 67 km fiber installed under a lake [12] and in a 96 km installed fiber [13]. A long time continuous QKD operation exceeding a few week period with an average secret key rate of 13 kbps was also demonstrated through a 16 km aerial fiber cable [14]. The plug and play scheme, however, suffers from backscattered light noise and may give an eavesdropper a chance of Trojan-horse attacks. In addition, complicated control of optical modulators for bidirectional transmission limits the key generation performance, such as a sifted key rate of 8.2 bps [13].

One-way transmission scheme with time-bin encoding was developed to solve these problems, and tested by a 125 km installed commercial fiber, although the key rate was still poor (not explicitly reported) [15]. The next challenge is to raise the key rate in a field environment for distances in the 100 km range.

To this end, several useful techniques have been developed to address the challenge. Stable optical interferometers based on PLCs allow one to realize long time stable operation. Some deficits of this technique such as the thermo-optic effect and polarization sensitivity were recently solved, that is, fine control of phase and polarization insensitivity could be simultaneously assured by temperature regulation with a precision of 0.01 K, attaining the total extinction ratio as high as 20 dB [17, 18]. In addition, use of SSPDs [19-21] drastically improved the QKD performances thanks to their low-noise- and high-speed-characteristics, and high stability [11].

*2.2 Requirement for stable clock synchronization*

Clock synchronization is crucial for a practical QKD system besides the improvement of the quantum transmission. The receiver will not identify the quantum signal correctly without the time gate synchronized to the photon arrival. The arrival time may drift due to the change in the transmission distance. For example, temperature variation of one degree results in more than one meter change in a 100 km fiber, which corresponds to 5 ns propagation delay in an optical fiber or three timeslots in a 625 MHz clock system. Because the clock synchronization from only quantum signals at the single photon level is quite challenging, it is much more convenient to transmit a classical clock signal, to compensate the fluctuation of the photon arrival time, in parallel with the quantum signal. One simple way is to use paired fibers. It is, however, not always easy to find an appropriate pair in existing installed fiber network infrastructure, and burdens the end users. Temperature variation may still induce a difference in the lengths of the paired fibers, resulting in degradation of the synchronization accuracy. These problems can be avoided by transmitting the clock and quantum signals through the same fiber with appropriate multiplexing.

There are two types of multiplexing: time-division multiplexing (TDM) and WDM. The former scheme requires an optical switch [22], a sufficiently long time interval between quantum and clock pulses, an extremely high extinction ratio for the clock pulses, and a precise compensation of the distorted waveform caused by chromatic dispersion to avoid interference between quantum and clock pulses. On the other hand, WDM synchronization [23-26] is free from those constraints that limit the key generation rate in TDM synchronization systems. However, the multiplexed clock signal will produce noise photons through the third order nonlinear effects, parametric amplification and spontaneous Raman emission in the fiber [27, 28]. These noise photons can easily reach a level comparable to the quantum signal level, especially for a long transmission distance such as 100 km, because the intensity of the clock signal is extremely larger by several tens of decibels than that of the quantum signal. If the nonlinear crosstalk is sufficiently suppressed, then the WDM scheme would be an ideal clock delivery scheme for a high speed, long distance, and long term QKD. One reported scheme to avoid this nonlinear crosstalk at 100 km-class QKD experiment is to allocate the wavelength of clock signal far away from that of the quantum signal, such as O- and C-band [24]. However, in order to increase transmission distance longer than 100 km, it is preferable to allocate these two signals in C- or L-band because of the high transmission fiber loss at O-band range.

The nonlinear crosstalk could be suppressed by using the WDM scheme depicted in Fig. 1. The optical power of the clock signal through the fiber was set as low as possible to suppress the unwanted noise generation, but at the same time high enough for the receiver. To this end we placed an optical amplifier right in front of the clock receiver. A narrowband optical filter (NBF) for the clock signal was necessary after the optical amplifier to eliminate the amplified spontaneous emission (ASE) noise. Another NBF for the quantum signal is also necessary to eliminate the noise photons that arise from the fiber nonlinear effects in a broad wavelength region. In order to reduce the noise photons further, the quantum signal is placed at the shorter wavelength side (1550 nm wavelength), i.e. anti-Stokes side of the clock signal (1570 nm wavelength), because anti-Stokes emission is weaker than Stokes emission by about 1.5 dB within ± 20 nm range of the pump source.

## 3. QKD field test

*3.1 Experimental setup*

A high speed QKD system for long distance transmission has been developed with the synchronization scheme described above combined with PLC based one-way interferometers and SSPDs for implementing the BB84 protocol. Figure 1 represents a schematic of our field test setup. We measured the transmission data, i.e., the sifted key rate and QBER to examine the long distance and high speed performance of the system.

In the transmitter, a 1550 nm directly modulated distributed feedback laser diode (DML) created 100 ps-wide pulses with a repetition rate of 625 MHz. The DML was driven with its DC bias under the lasing threshold to randomize the phase of each optical pulse, which reduces the possibility of Eve's use of the phase reference [29]. A two-input, two-output asymmetric Mach-Zehnder interferometer (2 × 2 AMZI) PLC split these pulses into a pair of coherent double pulses with an 800 ps time delay. A dual-drive MZ modulator (MZM) produced four quantum states in the time-bin encoding, according to two pseudorandom numbers made from a Pseudo Random Binary Sequence (PRBS) $2^7-1$. This pattern length is enough to investigate the transmission performance and the pattern dependence. The equatorial and polar states on the Bloch sphere, |0>, |0>-$i$|1>, |0>+$i$|1>, and |1>, were chosen for the time-bin encoding to remove optical modulators from the receiver [30]. The average photon number was set to estimate the final key rate by applying decoy state method [31, 32] using an optical attenuator (ATT) to 0, 0.05, 0.15, and 0.4 photons/pulse after the dispersion compensation fiber (DCF) prepared to mitigate the pulse broadening of the quantum signal. The quantum signal was combined with the clock signal by a high-isolation WDM coupler (isolation > 80 dB) [33], and was transmitted through the same fiber.

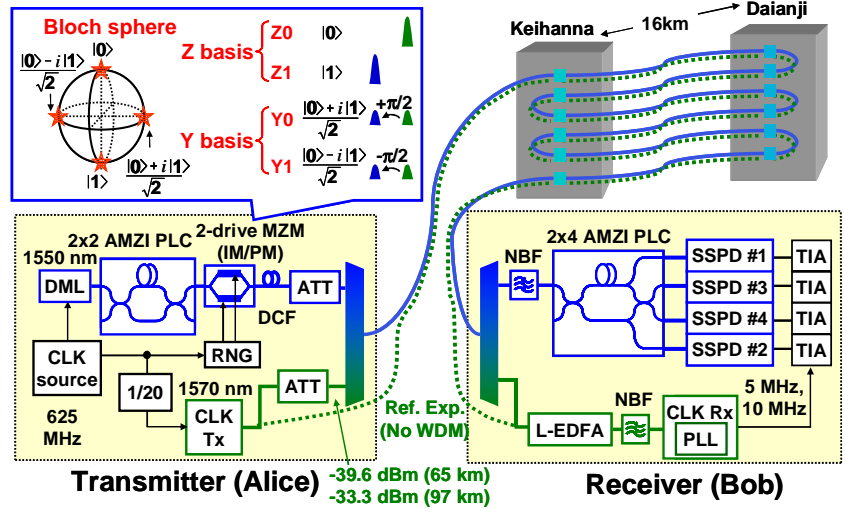

Fig. 1. QKD field test setup. The blue, green, and black lines indicate the route of the quantum signal, the clock signal, and the electric control signal, respectively. RNG, random number generator; ATT, optical attenuator; CLK source, clock source; CLK Tx, clock transmitter; CLK Rx, clock receiver. The balloon on the upper left corner explains the four quantum states used at this experiment. Relative phase and intensity of the coherent double pulses were adequately modulated using the 2-drive MZM.

The transmission fiber is an installed single-mode fiber (SMF) between Keihanna station at Kyoto and Daianji station at Nara in the National Institute of Information and Communications Technology (NICT) open test bed network, called the JGN II. The transmission loss was 13.2 and 20.2 dB for 65 km and 97 km, respectively. The dotted line in Fig. 1 represents a fiber for a reference experiment, where the clock signal was sent through a paired fiber, referred to as no-WDM. The difference in the lengths of the paired fibers was stable for a few hour duration.

In the receiver, the quantum signal and clock signal were divided by a WDM filter. The quantum signal was first filtered by a NBF (100 GHz full-width-at-half-maxima), and was discriminated by a two-input and four-output (2 × 4) custom AMZI PLC. Here it should be stressed that the decoding scheme is totally passive, reducing the burden of the decoder control [34]. Instead of 2 × 4 AMZI, it is more effective to decode the quantum signal by using a coupler sending the quantum signal either to a 2 × 2 AMZI or to single SSPD measuring the arrival time. This decode scheme requires only three detectors, however, the peripheral circuit comes to be more complex. In the present work, we employed 2 × 4 AMZI for sake of the simplicity. Four SSPDs based on niobium nitride nanowire, made by Moscow State Pedagogical University or NICT, and packaged by National Institute of Standards and Technology (NIST), received the quantum signal at each of the four outputs of the 2 × 4 PLC AMZI. The quantum efficiency and the dark count rate of these SSPDs were 1.2-1.6% and 90-160 count/s, respectively. Recently, gigahertz-class QKD experiments have been reported using InGaAs/InP avalanche photodiode [35-37], nevertheless SSPD is suitable for the present experiment because it shows extremely low dark count, is virtually free of afterpulsing, and is free from complex gate timing control. Four time interval analyzers (TIAs) recorded the photon arrival times with the reference clocks. After the recording at TIAs, meaningful detection events were extracted with a 400 ps-wide timing gate from each 1.6 ns time interval. The pulse width of the quantum signal (initially 100ps) was broadened to about 600 ps due to timing jitter of the whole system and waveform distortion caused by the residual chromatic dispersion. So about 7% of measured clicks occur outside the 400ps timing gate, and were actually discarded. This can, on the other hand, reduce the noise counts by a factor of 1/4.

The clock signal of 31.25 MHz at 1570 nm wavelength is amplified by an L-band erbium doped fiber amplifier (L-EDFA) with 25 dB gain, and passes through an NBF of 100 GHz full-width-at-half-maxima. The reference clocks for the TIAs (5 MHz and 10 MHz) are converted from the 31.25 MHz clock by using a phase locked loop (PLL) circuit. The optical power of the clock signal was set to -39.6 dBm for 65 km and -33.3 dBm for 97 km. As the isolation of the WDM coupler is higher than 80 dB, the intensity of stray light before transmission was suppressed to no higher than -110 dBm, which corresponds to $\mu_{stray} \sim 10^{-4}$ photons/pulse at Alice's transmitter. Nonlinear crosstalk arisen at the transmission line is the remaining problem to be revealed [27, 28], however, the numerical analysis of the phenomena is an issue in the future.

### 3.2 Key generation performance

Figure 2 shows the temporal fluctuation of the measured QBER (Fig. 2(a)) and the sifted key rate (Fig. 2(b)) as a function of the elapsed time for the WDM synchronization scheme. The average photon number was 0.4 photons/pulse. Both QBER and key rate were remarkably stable. (Although the stability should have been tested for a longer time, we could not do it this time only because of the limited time at the JGN II testbed.) The averaged QBERs were 1.36±0.05% and 2.89±0.07%, and the averaged sifted key rates were 11.51±0.32 and 2.40±0.06 kbps during 30 minutes and 80 minutes transmission for 65 km and 97 km, respectively. The error bars are the standard deviation in 12 and 22 trials for 65 km and 97 km transmission. Dark count contributions in the measured QBERs are calculated to be 0.3% and 1.3% for 65 km and 97 km, respectively, and remaining QBER components, about 1%, were attributed to the interferometer visibility and the room light directly injected to the SSPDs. Obtained sifted key rate, 2.40 kbps after 97 km transmission with the average photon number of 0.4 photons/pulse, is 100 times higher than the previous field experiment over 96 km [13], if the difference in the average photon number is taken into account.

Figure 3 shows the measured QBER (Fig. 3(a)) and the sifted key rate (Fig. 3(b)) versus average photon number for the WDM and no-WDM synchronization schemes. For both transmission distances 65 km and 97 km, the two schemes produced no significant difference in the QKD performances. This means that our WDM synchronization scheme succeeded in suppressing the crosstalk between the quantum and clock signals and the noise photons. After 97 km transmission, the averaged QBERs were 9.49±0.18% and 5.32±0.27%, and the averaged sifted key rates were 0.28±0.01 and 0.93±0.05 kbps for 0.05 and 0.15 photons/pulse, respectively. By applying a decoy state method analysis [31, 32] with an asymptotical estimation and Shannon-limited error correcting code, the upper bound of secure final key rate is estimated to be 0.78-0.82 kbps, this range comes from the difference of transmission performance between each basis, from above 97 km transmission data with three intensities: $\mu$ = 0.4, 0.15, and 0 photons/pulse.

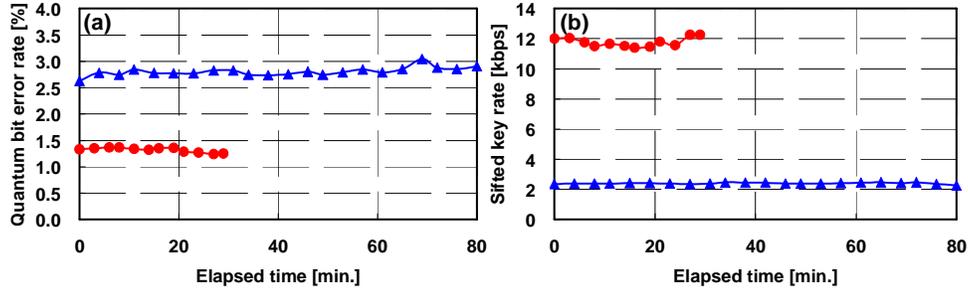

Fig. 2. Temporal fluctuation of the key generation performance. (a) Quantum bit error rate; (b) Sifted key rate. The data points are measured values for 65 km (red circles) and 97 km transmission (blue triangles).

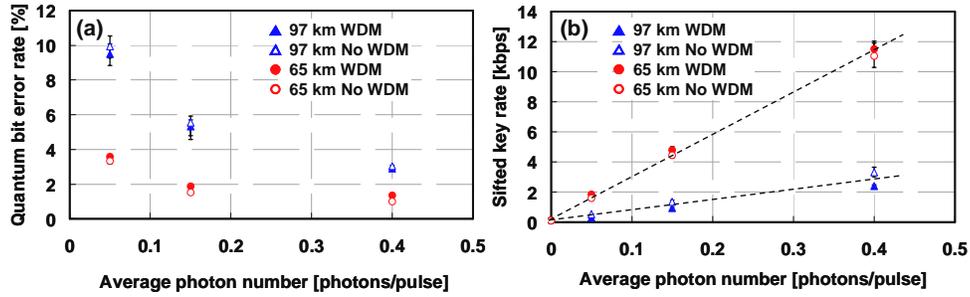

Fig. 3. Key generation performance against average photon number. (a) Quantum bit error rate; (b) Sifted key rate. The data points are measured values for 65 km transmission (red circles) and 97 km transmission (blue triangles). Solid plots show the measured values obtained from the WDM synchronization and hollow plots show those obtained from the no-WDM experiments. The error bars show the standard deviation.

## 4. Conclusion

We performed a QKD field test through a 97 km installed fiber using PLC based one-way interferometers, SSPDs, and WDM synchronization. We established the clock transmission technique without degrading the quantum signal. We demonstrated BB84 key generation in a field fiber stable over an hour with the fastest-ever sifted key rate of 2.4 kbps and QBER of 2.9% after 97 km transmission. The sifted key rate can reach a hundred kbps, which corresponds to 30 kbps of the secure key rate, in the near future by improving the system detection efficiency using SSPDs to 10% and increasing clock frequency to 3 GHz.

The present experiment was exhibited in order to confirm a performance of a high speed and long distance QKD transmission and WDM clock synchronization in the field environment, hence obtained key was not perfectly secure. A full QKD system with a security guarantee can be obtained by using a true random binary sequence in place of PRBS, selecting the average photon number randomly for each pulse to apply decoy method, building up a high-speed data processing system including a huge amount of memory for random number storing, and distilling the final key with a protocol for finite length codes, as described in Ref. [38, 39]. A precise calibration should be also necessary to satisfy the assumption in the security proof [31, 32]; the four detectors should be controlled to show the identical performances. The reported QKD transmission system is compatible to the above additional functions, and will serve as a solid platform for the complete QKD systems.

## Acknowledgements

A part of this work was supported by the project on "Research and development for practical realization of quantum cryptography" of NICT under the Ministry of Internal Affairs and Communications of Japan, and NIST Quantum Information initiative program.